\documentclass[fleqn,12pt]{wlpeerj}

\usepackage[colorlinks,citecolor=blue,urlcolor=blue]{hyperref}
\usepackage{float}
\usepackage{geometry}
\title{Spatio-temporal modeling of co-dynamics of smallpox, measles and pertussis in pre-healthcare Finland}

\author[1]{Tiia-Maria Pasanen}
\author[1,2]{Jouni Helske}
\author[1]{Harri Högmander}
\author[3,4]{Tarmo Ketola}
\affil[1]{Department of Mathematics and Statistics, University of Jyväskylä}
\affil[2]{INVEST Research Flagship Centre, University of Turku}
\affil[3]{Department of Forestry, University of Helsinki}
\affil[4]{Department of Biological and Environmental Science, University of Jyväskylä}
\corrauthor[1]{Tiia-Maria Pasanen}{tiia-maria.h.pasanen@jyu.fi}

\begin{abstract}
Infections are known to interact as previous infections may have an effect on risk of succumbing to a new infection. The co-dynamics can be mediated by immunosuppression or -modulation, shared environmental or climatic drivers, or competition for susceptible hosts. Research and statistical methods in epidemiology often concentrate on large pooled datasets, or high quality data from cities, leaving rural areas underrepresented in literature. Data considering rural populations are typically sparse and scarce, especially in the case of historical data sources, which may introduce considerable methodological challenges. In order to overcome many obstacles due to such data, we present a general Bayesian spatio-temporal model for disease co-dynamics. Applying the proposed model on historical (1820--1850) Finnish parish register data, we study the spread of infectious diseases in pre-healthcare Finland. We observe that measles, pertussis, and smallpox exhibit positively correlated dynamics, which could be attributed to immunosuppressive effects or, for example, the general weakening of the population due to recurring infections or poor nutritional conditions.
\end{abstract}

\begin{document}

\flushbottom
\maketitle
\thispagestyle{empty}

\section*{Introduction}

Infections exist rarely in isolation and their effects on hosts are known to interact and to have both positive and negative relationships between each other \citep{gupta1998, rohani2003, shrestha2013, mina2015, nickbakhsh2019}. For example cross immunity may prevent others infecting the host, competition for same resources or susceptible host can have strong effects on epidemics, and sometimes one infection paves a way for another \citep{gupta1998, rohani2003, graham2008}. Perhaps historically the best-known relationship between infections is the immunosuppressive effect of measles on the following pertussis epidemic by increasing the severity of the epidemic \citep[see][]{coleman2015, mina2015, noori2019}. Coinfections of parasites \citep{graham2008} and viruses and respiratory bacterial infections are well known \citep[e.g.,][]{bakaletz2017, wong2023}, whereas understanding coinfections and cotransmissions of, for example, zika, dengue and chikungunya viruses presents a current serious challenge for public health \citep{vogels2019}.

Demographic consequences of epidemics are most dramatically seen in large cities and in densely populated areas, which is reflected in the epidemiological research in general \citep{mueller2020}. However, as rural areas constitute a large part of most of the countries, the spatio-temporal dynamics of epidemics in populations with low densities deserve more attention \citep{mueller2020}. In rural areas populations often consist of loosely connected metapopulations rather than large and epidemiologically more autonomous populations in cities. This has most likely strong repercussions to the drivers of epidemics \citep{ball2015} and also to the co-occurrence of infections. However, these issues are rarely addressed in literature, possibly due to the statistical challenges encountered with sparse and scarce data, as well as the difficulty of modeling the dynamics of several infections simultaneously both in space and time.

The discrepancy between studying dense and sparse populations is evident and can be seen, for example, by comparing our case of rural Finland in 1820--1850 to the seminal research of \citet{rohani2003}. Their study is based on five large European cities, where the weekly number of fatalities frequently exceeds $30$ and even $80$. In our data, the recorded incidents in most of the towns rarely exceed one person per month, as we study a small and mainly agrarian population in the southern part of Finland with ca. $1.2$ to $1.6$ million individuals \citep{voutilainen2020}. The population, without proper healthcare \citep{saarivirta2012}, was spread over a vast area in geographically separated, but socially connected, small towns and villages. Based on the data from 1882, population sizes of towns varied between $300$ and $25{,}000$ \citep{suomenmaan1882, ketola2021}. Despite the obvious uncertainty of population censuses during that era \citep{voutilainen2020}, the contrast between our data and most of the published datasets is striking. Statistical modeling of such data is problematic due to incomplete information from some locations and the rare occurrence of events, hampering the ability of generally used models to consider several infectious diseases at the same time and on both temporal and spatial scales.

To estimate the spatio-temporal co-dynamics of deaths due to pertussis, measles and smallpox, we build a model that can overcome the limitations inherent in our data. The model jointly estimates the spread of multiple infections, enabling the exploration of the temporal and spatial dependence structures both within and between the infections. Our general Bayesian model consists of a multivariate latent incidence process, a seasonal component, and multiple predictors whose effects may vary between the towns. This allows us to study the dynamics of the diseases simultaneously despite having only incomplete information about the deaths. The results we get from modeling the mere presence-absence data are compared with those of modeling the death counts, and the simplification is deemed to be a reasonable option in our case. Given the limitations of our data, we do not aim to make causal claims on a biological level, and rather than focusing on the magnitude of infections and the intensity of deaths, our primary interest lies in understanding how the prevalence of these three diseases varied both spatially and temporally in pre-industrial Finland, and if there were possible associations in these dynamics across the diseases.\footnote{This text was originally published as a preprint (https://export.arxiv.org/abs/2310.06538).}

\section*{Materials and methods}

\subsection*{Data}

During the study period, 1820--1850, the parishes in Finland kept track, among others, of baptisms, burials and causes of deaths, according to common and long held principles \citep{pitkanen1977}. Even though the death diagnostics were based on symptoms, some infections can be considered to be diagnosed rather accurately due to their characteristic features. These diseases include pertussis (whooping cough), measles and smallpox, which we consider. These infections were the main reason for child mortality, and, overall, they were responsible for approximately $5$, $3$, and $3$ percent of total deaths, respectively, according to our data. Based also on the available records in our data, the median ages of deaths in complete years were $0$ (sd = $3.6$) for pertussis, $2$ (sd = $3.9$) for measles, and $2$ (sd = $7.5$) years for smallpox. Smallpox vaccinations were started in Finland in 1802 and were slowly progressing during the study period \citep{briga2022}. However, general healthcare was almost non-existent as in 1820 there were only $373$ hospital sickbeds for $1.2$ million inhabitants \citep{saarivirta2012}.

Our data consist of the daily numbers of deaths, classified by the cause of death, between January 1820 and December 1850 from $N = 387$ different regions (towns) in mainland Finland with the exclusion of northern areas. The time window is chosen such that there were no major famines, wars, border changes, or other potentially confounding events, which could have altered the geographical partition or the dynamics of the epidemics. The general stability achieved is advantageous in the modeling.

\begin{figure}[]
    \centering
    \includegraphics{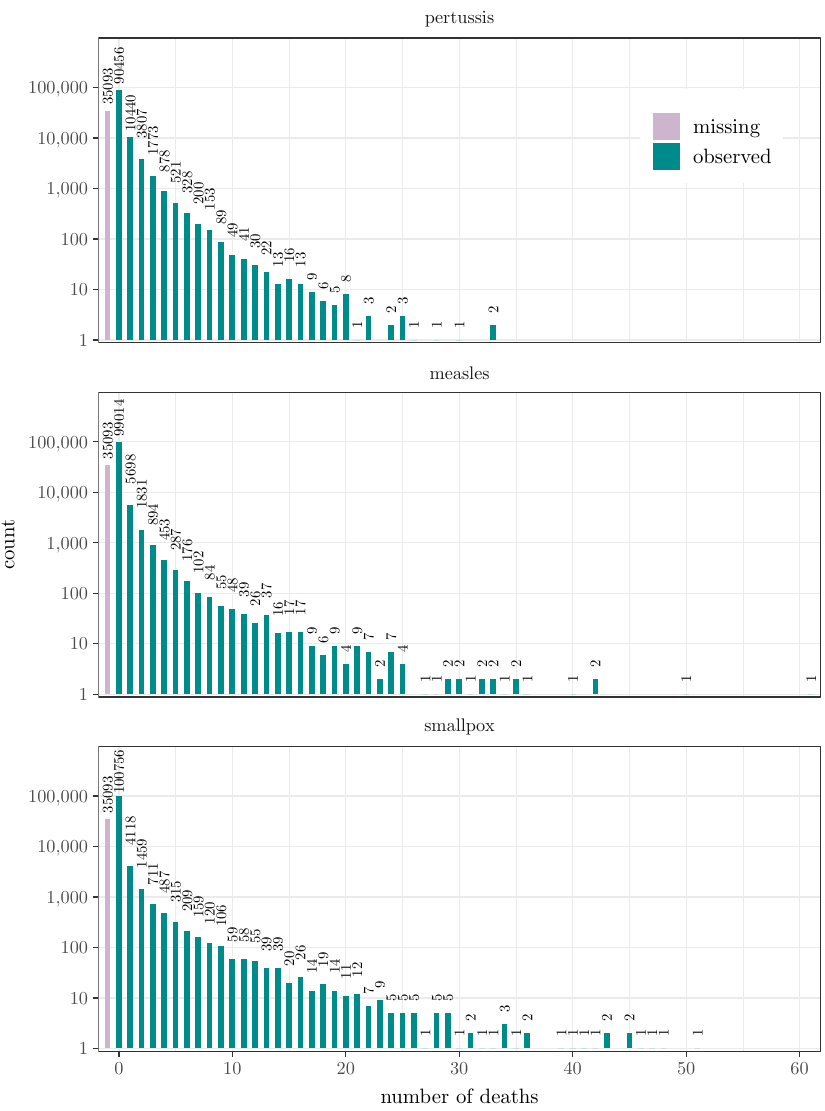}
    \caption{The counts of the observed numbers of deaths over all towns and months plotted by disease. The first bar indicates the number of missing observations. The numbers above the bars are the counts. Note that the y axis is on logarithmic scale.}
    \label{fig:counts}
\end{figure}

Although using the daily time scale would, in theory, be ideal for modeling disease dynamics, the infrequency of deaths implies that our data lack sufficient information to study temporal dependencies at such a detailed level. This issue is likely further compounded by the spatial heterogeneity of the data and the potentially complex lagged auto- and cross-dependencies between the infection dynamics of the diseases. Therefore, the daily counts of deaths are aggregated over time into a monthly level, decreasing the number of zero observations yet maintaining a reasonable time resolution for observing the spread of the diseases on our geographical scale. This yields a total of $T = 372$ time points. The counts of the observed numbers of deaths by disease considering the aggregated data are visualized in \autoref{fig:counts}. In each case of the three infections, about $93-96$\% of the death counts are zero or one, and less than $2$\% of the counts are more than three, despite the aggregation. The reliability of the actual death counts varies considerably both temporally and spatially owing to the heterogeneous quality of the parish records and the cause of death classifications. Moreover, these deficiencies are not necessarily independent of the true number of deaths. It is also noteworthy that despite the records of baptisms and burials there are no reliable estimates of the population sizes at the town level \citep{voutilainen2020}. Hence the intuitive idea of using local relative mortality is unfortunately beyond reach. Because of this and the aforementioned reliability issues---and since most of the counts are still either zero or one even after the aggregation to a monthly level---only the dichotomous knowledge of the death occurrence is used in the main analysis. Even so, the count data are considered in the model comparison section.

About $24\%$ of the data concerning death occurrences in a town and a month are missing, and from $57$ out of the overall $387$ regions there are no observations at all. The missingness pattern is common to all diseases, i.e., for a particular town and month, we have observed counts for all diseases or for none of them. This is because the missing data can be attributed to the absence of parish records that document all deaths. This could be a result of incomplete digitalization of the parish records, loss of such documents for example due to fire in rectory or church, or simply because there were no deaths in a given month. Therefore, the missing data could potentially depend on the unknown population sizes, but not on specific causes of death, given that the proportions of deaths attributable to the particular cause are relatively small. Nevertheless, our proposed model provides estimates of the monthly probabilities of observing at least one death also for these towns.

\subsection*{Model}

We construct a general model to describe the spatial and temporal dependencies both within and between the infections under study. We also want to enable exploiting other relevant information as explanatory variables. In epidemiological context there typically occur spatial or temporal trends or seasonal effects, which can be included as separate components in the model. Due to the nature of our data, we model the probability of observing at least one death caused by a disease in a certain town in a certain month. The model consists of a trend, a seasonal effect, and a regression part reflecting the local effects of the previous state of infection in the focal town and its neighboring towns.

Formally, let $y_{i, t}^d$ denote a dichotomous variable of an event where at least one death occurs due to a disease $d$ in a region $i = 1,\, \dots,\, N$ at a time point $t = 1,\, \dots,\, T$, where $N$ is the number of regions and $T$ the number of time points. Let $K_x^d$ indicate the number of explanatory variables $x$, based on the features of the region $i$. Accordingly, $z$ denotes the explanatory variables, and $K_z^d$ their number, related to the neighborhood of the region $i$. Thus the model for observing at least one death caused by the disease $d$ in the region $i$ at the time point $t$ can be written as follows:

\begin{equation}\label{eq:model}
    y_{i,t}^d \sim \text{Bernoulli}\!\left(\!\text{logit}^{-1}\!\left(\eta_{i,t}^d\!\right)\!\!\right)\!\!,
\end{equation}
where
\begin{equation}\label{eq:eta}
    \eta_{i,t}^d = \lambda_{i}^d \tau_{t}^d + s_t^d + a_i^d + b_i^d \sum_{k = 1}^{K_x^d} \beta_{k}^d x_{i, k}^d + c_i^d \sum_{k = 1}^{K_z^d} \gamma_{k}^d z_{i, k}^d.
\end{equation}
Here Bernoulli distribution with a logit link is chosen due to our dichotomous consideration of the occurrences of death, but other distributions with appropriate link functions can be applied for different types of response variables.

The first three terms being summed in \autoref{eq:eta} form a base level, in our case, for the probability of observing at least one death caused by the disease $d$ at each town and month. More specifically, the first term consists of the time dependent latent factor $\tau_t$, describing the nationwide incidence (on log-odds scale), or trend, of the disease $d$, and the regional adjustments or loadings $\lambda_i$, with respect to the mean level. As in general dynamic factor models, the products $\lambda_i \tau_t$ are not identifiable without constraints \citep{bai2015}. Instead of the common approach of fixing one of the loadings $\lambda_i$ to 1, we constrain the mean of the loadings to $1$, enabling the interpretation of the factor $\tau_t$ as the nationwide incidence level. Due to the nature of the other terms in \autoref{eq:eta}, this incidence level gives the national average log-odds of observing at least one death in an "average" town in a given month if no deaths were observed in the previous month in the focal town or in its neighboring towns. The second term $s_t$ is a monthly seasonal effect, which is the average deviation from the nationwide incidence level, summing up to $0$ over the months. The third term $a_i$ is a regional, zero-mean constant reflecting local deviations from the nationwide incidence level $\tau_t$ due to unobserved local demographic, geographic, social or other characteristics associated with mortality.

The last two sum terms in \autoref{eq:eta} form the regression part of the model. The first sum includes the covariates $x_{i,k}$ related to the focal region $i$, and the second sum the covariates $z_{i,k}$ related to the neighboring regions. These variables have both nationwide coefficients $\beta$ and $\gamma$, and their local adjustments $b_i$ and $c_i$ amplifying or diminishing the nationwide level. As the regional constants $a_i$, also the multiplicative local adjustments may account for any unobserved heterogeneity between the regions, such as the local population sizes or densities. The adjustment parameter $b_i$ reflects the features of the focal region $i$, and $c_i$ those of the neighborhood of the region $i$ (possibly relative to $i$). We assume that $b_i$ is same for all covariates $x_k$, and, accordingly, $c_i$ for all $z_k$, since the underlying regional characteristics modifying the nationwide mortality effects $\beta$ and $\gamma$ do not depend on the covariates.

In our study there are three covariates assigned to the town $i$ and another three to its neighbors for all diseases $d = p,\, m,\, s$, where $p$ stands for pertussis, $m$ measles and $s$ smallpox. The local explanatory variables $x_{i,k}$ are the presences of deaths caused by the three different diseases in the previous month in the focal town, whereas the regional neighborhood predictors $z_{i,k}$ are the averages of the same death presences over the local neighborhood. We define two regions being neighbors when they share a border. By this definition, all the towns have at least one neighbor. Other definitions of neighborhood could be used as well, for example based on the transportation network or distance, leading to weighted averages of death occurrences.

As noted earlier, our data contain a large number of missing observations. We assume that the probability of having a missing response or predictor variable is independent of the value of the response variable. This presumption may be considered valid, given that the lack of parish records on deaths is unlikely to depend on the causes of deaths in a particular month. Under this assumption, we can model the observed data analogously to a complete-case analysis in an unbiased manner, assuming our model is correctly specified \citep{VanBuuren2018}. This eliminates the need for multiple imputation or sampling missing observations using MCMC algorithms which do not use gradient information (i.e., algorithms capable of sampling discrete variables), both of which would be computationally unfeasible in our Bayesian spatio-temporal context. 

In practice, the complete-case analysis in our context means that to use an observation as a response, we require that both the current response variable as well as all of the related covariates are observed. If any of them is missing, we omit the particular combination of town and month as a response. On the other hand, when calculating the neighborhood covariates $z_{i,k}$, which in our case are the averages over the observations within the neighborhood, we omit the neighbors with missing observations so that they are not included even as a denominator in the evaluation of the mean. If all neighbor observations are missing, the corresponding covariate is defined as missing.

We model the latent factor $\tau_t = (\tau_t^p,\, \tau_t^m,\, \tau_t^s)$, the temporal process describing the baseline of the nationwide incidence rates, as an intercorrelated random walk, $\tau_{t+1} \sim \text{N}(\tau_t, \Sigma)$. Here $\Sigma$ is an unconstrained $3 \times 3$ covariance matrix parametrised using the standard deviations $\sigma_\tau^d$ and the correlation matrix $R$. In our application this latent process, together with the regional constant and the monthly effect, can be interpreted as the probability to observe at least one new death in a particular town when no deaths caused by any of the three diseases were observed in the previous month in the focal town or in its neighborhood.

In the Bayesian modeling framework we need prior distributions for all the parameters to be estimated. The incidence factors $\tau^d$ follow a N($-2, 2^2$) prior at the first time point and form a random walk at later time points. For the correlation matrix $R$ we use LKJ(1) prior with Cholesky parametrisation \citep{lewandowski2009}, i.e., uniform prior over valid $3\times 3$ correlation matrices. For the regional parameters we apply Gaussian priors: $\lambda_i \sim \text{N}(1,\, \sigma_{\lambda^d}^2),\, a_i \sim \text{N}(0,\, \sigma_{a^d}^2),\, b_i \sim \text{N}(1,\, \sigma_{b^d}^2)$, and $c_i \sim \text{N}(1,\, \sigma_{c^d}^2)$. The prior means of the local adjustments $b_i$ and $c_i$ are set to 1, which enables us to interpret $\beta$ and $\gamma$ as nationwide effects as is the case with additive multilevel models with population-level and group-level effects. While we use a hard equality constraint for the mean of the $\lambda_i$s to ensure the identifiability and more efficient estimation of the model, the hierarchical priors for $b$ and $c$ are sufficient for their identifiability. For the unknown deviations $\sigma_{\tau^d}$, $\sigma_{\lambda^d}$, $\sigma_{a^d}$, $\sigma_{b^d}$, and $\sigma_{c^d}$ we assign Gamma priors with shape parameter $2$ and rate parameter $1$.  The nationwide coefficients $\beta$ and $\gamma$ have N($0,\, 2^2$) priors. The seasonal effects $s_t^d$ follow a standard normal prior with the aforementioned sum-to-zero constraint. The priors can be seen as weakly informative, and they are chosen primarily to enhance the computational efficiency \citep{banner2020}. 
 
Naturally, any of the components in the model could be excluded by setting the corresponding coefficients or standard deviations to zero. Our Bayesian model encompasses all such simplified alternatives, with the corresponding model and parameter uncertainty reflected by the estimated posterior distributions, leading to more truthful uncertainty estimates compared to merely imposing prior constraints on certain effects to be zero.

\section*{Results}

The model is estimated using Markov chain Monte Carlo (MCMC) with \textit{cmdstanr} \citep{gabry2022}, which is an R interface \citep{r} for the probabilistic programming language Stan for statistical inference \citep{stan}. To draw the posterior samples we use NUTS sampler \citep{hoffman2014, betancourt2018} with four chains, each consisting of $7{,}500$ iterations, the first $2{,}500$ of which discarded as warm-up. With parallel chains the computation takes about ten hours. The model is estimated on a supercomputer node with four cores of Xeon Gold 6230 $2.1$ GHz processors and $40$ GB of RAM. According to the MCMC diagnostics of the \textit{cmdstanr} \citep{vehtari2021} the model converges without divergences, the $\widehat{R}$ statistics are always below $1.005$, and the effective sample sizes are approximately between $700$ and $43{,}000$. The lowest effective sample size is the one of the deviation parameter of the constants considering measles, $\sigma_{\alpha_m}$. The R and Stan codes, the data used for the analysis, and Supplementary Figures and Tables are available on GitHub (\url{https://github.com/tihepasa/infectionDynamics}). All the figures were created using the R packages \textit{ggplot2} \citep{ggplot2} and \textit{ggpubr} \citep{ggpubr}.

\begin{figure}[]
    \centering
    \includegraphics{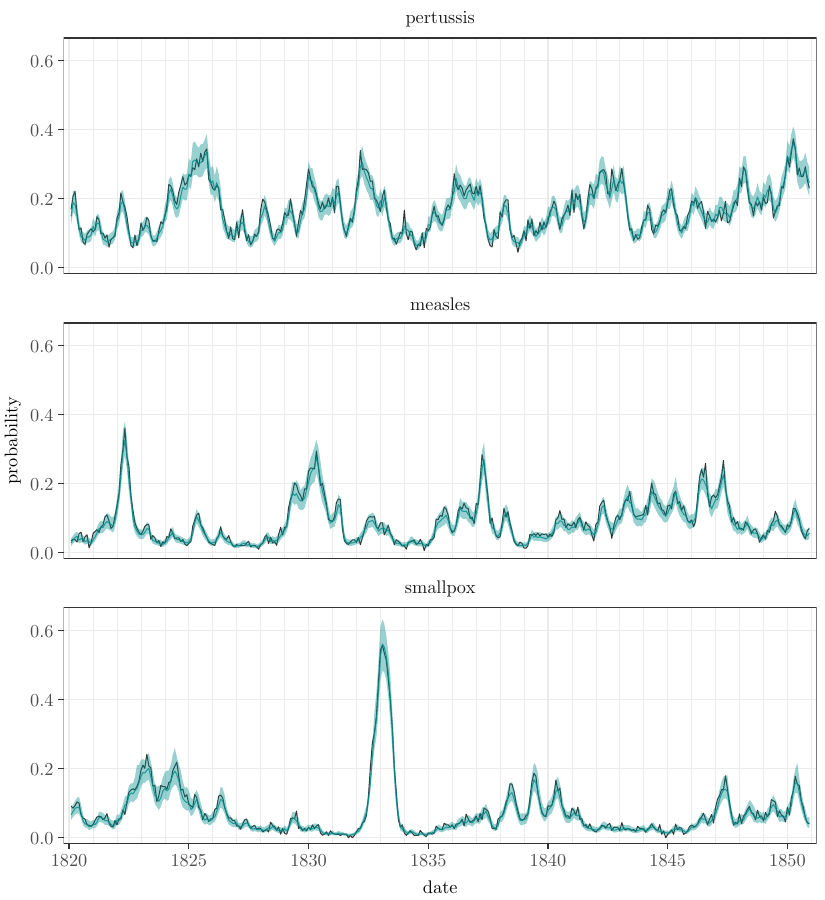}
    \caption{The prevalence of the diseases is illustrated with the dark lines depicting the proportions of the towns where at least one death was observed. The lighter turquoise lines show the posterior means of the corresponding estimates and the shaded areas their $95$\% posterior intervals. Note that the data line is calculated over the observed sites, whereas the estimate line averages all the sites.}
    \label{fig:series}
\end{figure}

\begin{figure}[]
    \centering
    \includegraphics{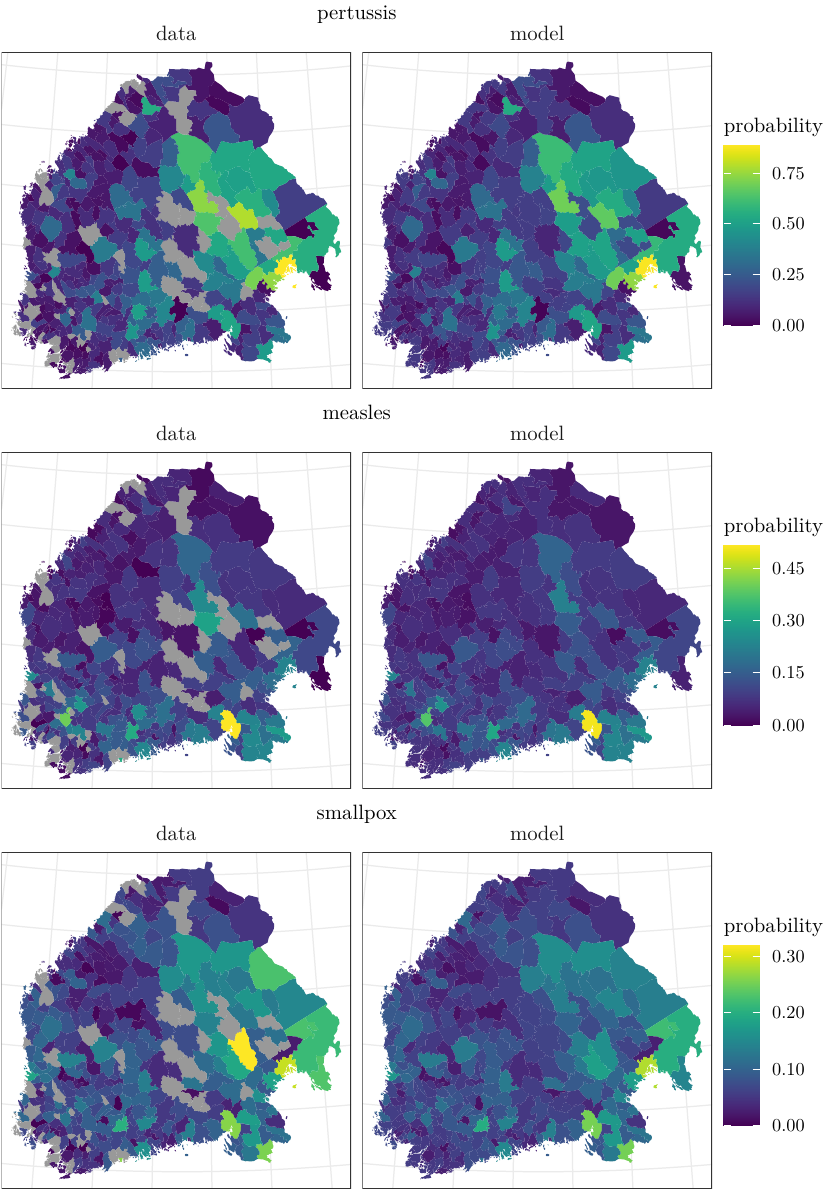}
    \caption{The left panels show the proportions of the months when deaths were recorded over the study period of 31 years. The gray areas indicate the regions where all data are missing. The right panels present the regional averages of the predicted conditional probabilities to observe at least one death caused by each disease in each month given the actual observations from the previous month. Note that the data are averaged over the observed towns and months, whereas the model covers all the sites and times.}
    \label{fig:maps}
\end{figure}

To visualize the temporal and spatial patterns of the death occurrences and to see how the model estimates the corresponding probabilities to observe at least one death, the data and the predictions based on the model are plotted as time series and as maps in Figures \ref{fig:series} and \ref{fig:maps}. The estimates are computed as $k$-step forward predictions, where $k - 1$ is the number of preceding missing months, conditional on the posterior distribution of model parameters, including the latent incidence process $\tau$. For the first time point in this calculation we also assume that the missing observations are zeros in order to have covariates for all the sites.

The temporal behavior of the data and the corresponding estimates are similar. The slight differences may be due to the fact that the proportions are based only on the data available, whereas the model predictions cover all the regions. The spatial patterns of the modeled probabilities reflect the infection distributions visible in the data. Pertussis, measles and smallpox all have emphasis on the eastern half of Finland, with especially measles extending its prevalence to the southern parts of the country as well. When it comes to the completely missing sites, the medians of the estimated average probabilities over time are $3.7$ percentage units higher for them than for the sites with observed data in case of pertussis, $0.3$ percentage units higher in case of measles, and $0.2$ percentage units lower in case of smallpox. The differences are quite small, and by the prediction account, the model seems to work well.

In what follows, we present the results in detail. They confirm that all the components in the model are relevant, capturing different aspects of the spatio-temporal dynamics of the epidemiological phenomena.

\begin{figure}[]
    \centering
    \includegraphics{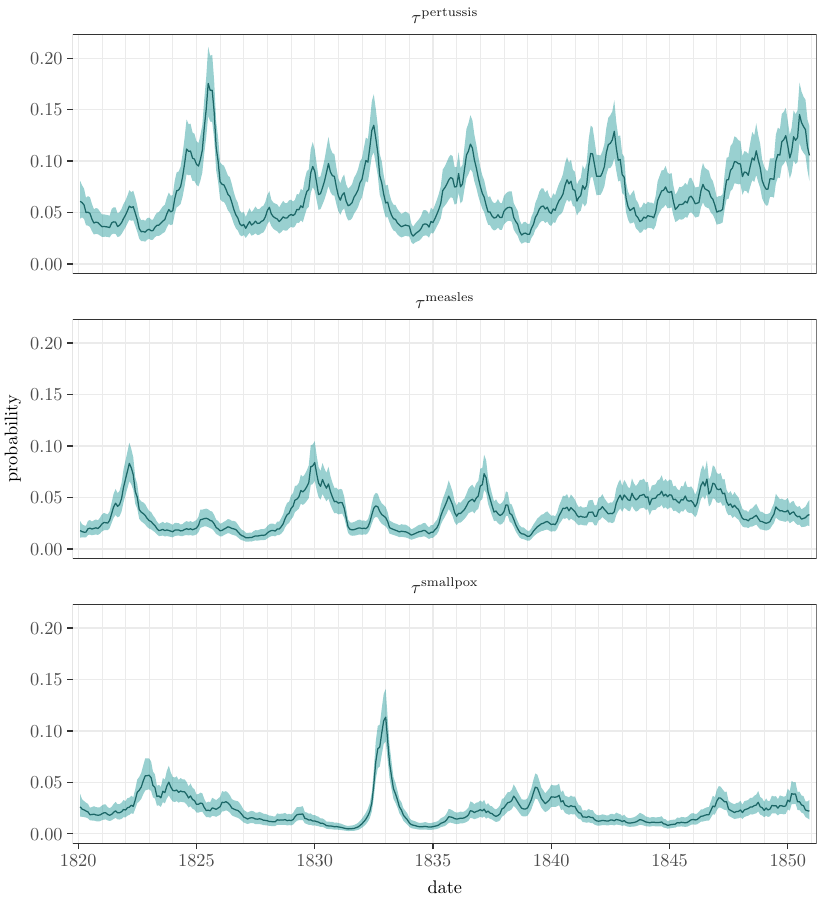}
    \caption{Posterior means and $95$\% posterior intervals for the unobserved incidence factors $\tau_{t}^d$ for pertussis, measles and smallpox over the time period under study. The curves are on a probability scale.}
    \label{fig:tau}
\end{figure}

The nationwide incidence time series of the diseases are depicted by the factors $\tau_t^d$. The corresponding estimates are shown in \autoref{fig:tau} on a probability scale ($\textrm{logit}^{-1}(\tau_t^d)$). In general, they seem to have the same shapes as the observed nationwide monthly proportions of towns where at least one death caused by pertussis, measles or smallpox was recorded (see \autoref{fig:series}). There is one major disease outbreak regarding smallpox, whereas the other diseases have several peaks, pertussis varying the most. No clear periodicity can be seen in any of the series, which was also confirmed by estimating dominant frequencies via spectral analysis using the R package \textit{forecast} \citep{Hyndman2008}.

\begin{figure}[]
    \centering
    \includegraphics{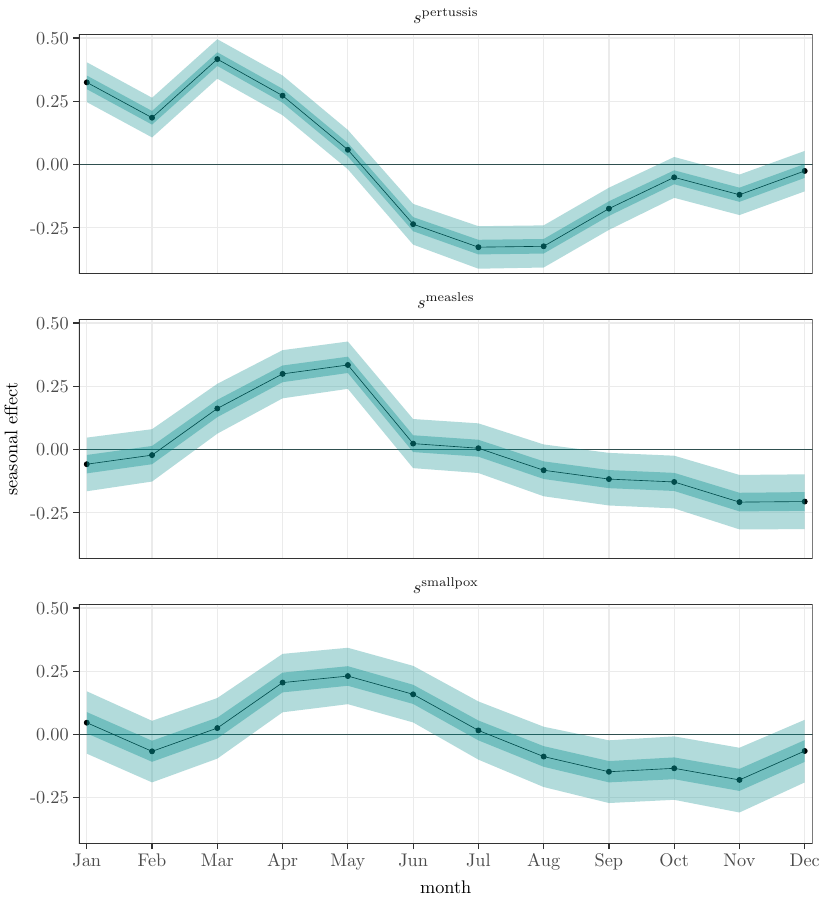}
    \caption{Posterior means (black), $50$\% (dark turquoise) and $95$\% (light turquoise) posterior intervals of monthly seasonal effects $s_t^d$ for pertussis, measles and smallpox over a year.}
    \label{fig:season}
\end{figure}

The seasonal effects $s_t^d$, or the average monthly deviations from the nationwide incidence level, are shown in \autoref{fig:season}. According to the estimates, the seasonal effect of pertussis peaks at the beginning of the calendar year, while the effect decreases during the summer and increases again towards the end of the year. In contrast, the only distinctive seasonal effects related to measles and smallpox are the peaks in the spring and the minor decreases at the end of the year.

Measured by the $\tau$ factors, we found a distinctive correlation between the infections of measles and pertussis, $0.33$ with a $95$\% posterior interval $[0.08,\, 0.55]$. Omitting the specific seasonal term $s$ in the model yields almost the same correlation $0.31$ $[0.10,\, 0.50]$. The correlation between smallpox and measles is ambiguous, being $0.24$ $[-0.01,\, 0.46]$, though it increases to $0.46$ $[0.26,\, 0.63]$ in the model without the seasonal component. This implies that monthly effects explain partly but not exhaustively the connection between these diseases. Smallpox and pertussis seem to be mutually independent, $0.06$ $[-0.19,\, 0.30]$, which is also the case with the model without the seasonal terms, $0.15$ $[-0.07,\, 0.37]$.

\begin{figure}[]
    \centering
    \includegraphics{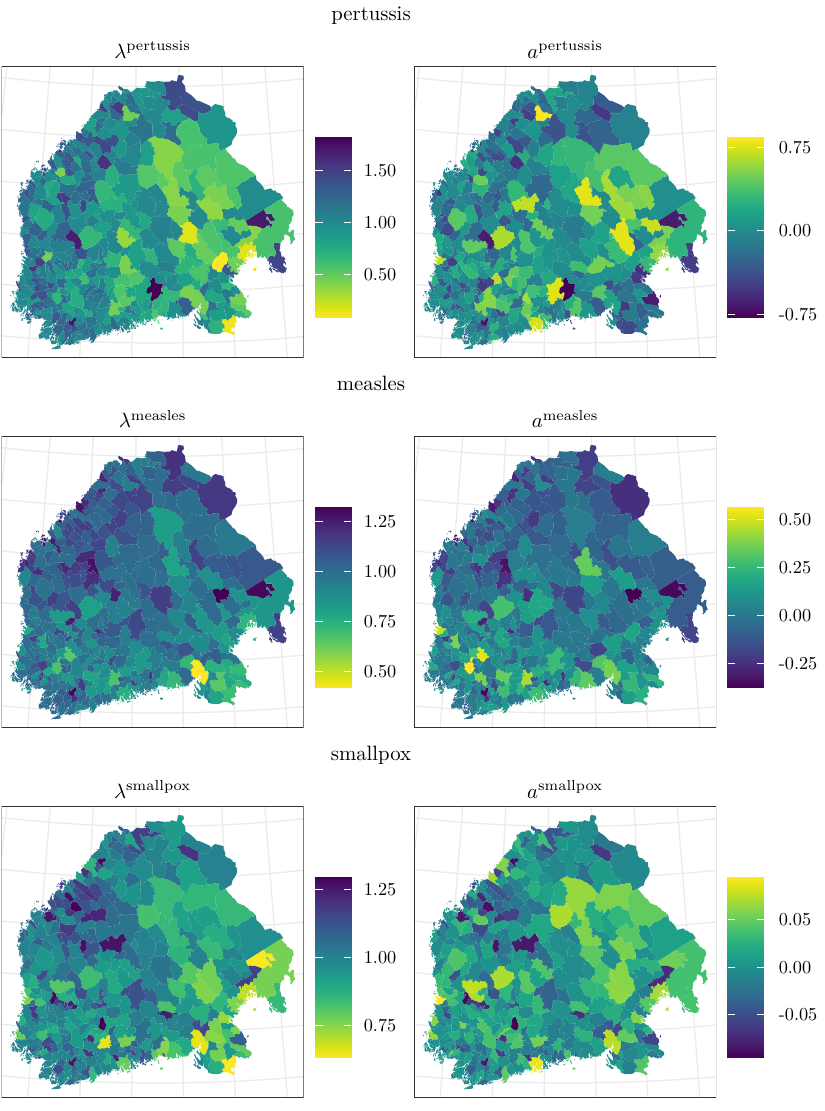}
    \caption{The left panels show the posterior means of the local loadings $\lambda_i$, adjusting the national incidence factors $\tau_t$. Since the factors are negative, the smaller the loading is, the greater the probability of at least one death is. The right panels illustrate the posterior means of the regional constants $a_i$.}
    \label{fig:lambda}
\end{figure}

According to the regional loadings $\lambda_i^d$ adjusting the nationwide factors $\tau_t^d$, it was more likely to die of any of these diseases in eastern and southeastern Finland than in other parts of the study area. This is also in accordance with the maps of the data in \autoref{fig:maps}. The posterior means of the loadings $\lambda$ are plotted in \autoref{fig:lambda}. Considering the loadings, there is most local variation in pertussis, $\sigma_{\lambda^p} = 0.35$ with a $95$\% posterior interval $[0.31, 0.40]$. With regard to measles and smallpox, the loadings vary less, $\sigma_{\lambda^m} = 0.17$ $[0.14, 0.20]$ and $\sigma_{\lambda^s} = 0.15$ $[0.13, 0.17]$.

The final term affecting the base level of the probability to observe at least one death caused by pertussis, measles or smallpox consists of the regional constants $a_i$, shown in \autoref{fig:lambda}. Those related to pertussis and smallpox seem to be larger in eastern and southwestern inland areas, whereas those considering measles are largest in southern Finland.

The estimates of the nationwide regression coefficients are represented in \autoref{tab:coefs}. All effects differing from zero are positive, meaning they increase the probability to detect at least one death. The probability to observe one or more deaths induced by one of these diseases is increased most prominently if there are recorded deaths caused by the same disease in the same town, or in its neighbors, in the previous month. However, there is more uncertainty in the effects of neighbors than in those of the towns themselves. The risk that there is at least one death caused by pertussis is increased by the occurrence of measles, whereas the corresponding effect of smallpox is not distinctive. Measles is probably affected more by smallpox than by pertussis. In turn, measles seems to affect smallpox more than pertussis does.

When it comes to the local adjustments $b_i$ and $c_i$, their standard deviations are clearly above zero, varying between $0.22$ ($\sigma_{b^m}$) and $0.55$ ($\sigma_{c^p}$), which indicates that the local adjustments differ around the country. There are no obvious interpretations of their spatial patterns (see maps of $b$ and $c$ in Supplementary Figures 1 and 2 on GitHub). This is credible, since the coefficients represent the combined effects of multiple unobserved features that are not necessarily spatially organized.

\begin{table}[]
\small
\renewcommand{\arraystretch}{1.2}
\centering
\caption{\label{tab:coefs}Posterior means and 95\% posterior intervals of the nationwide regression parameters grouped by the response disease. Superscript indicates the response disease and subscript the explanatory disease.}
%\vspace{0.3cm}
\begin{tabular}{l@{\hskip .8cm}lrr@{\hskip .8cm}lrr}
  \hline
  & \multicolumn{3}{l}{\hspace{-0.2cm}within towns} & \multicolumn{3}{l}{\hspace{-0.2cm}between towns} \\
  & & mean & (2.5, 97.5\%) & & mean & (2.5, 97.5\%) \\
  \hline
  pertussis $\rightarrow$ pertussis & $\beta_{p}^p$ & 1.56 & (1.47, 1.64) & $\gamma_{p}^p$ & 1.23 & (1.10, 1.35) \\
  measles $\rightarrow$ measles & $\beta_{m}^m$ & 1.90 & (1.82, 1.98) & $\gamma_{m}^m$ & 2.21 & (2.06, 2.38) \\
  smallpox $\rightarrow$ smallpox & $\beta_{s}^s$ & 2.43 & (2.34, 2.53) & $\gamma_{s}^s$ & 2.57 & (2.40, 2.74)\vspace{0.2cm} \\
 
  measles $\rightarrow$ pertussis & $\beta_{m}^p$ & 0.11 & (0.04, 0.18) & $\gamma_{m}^p$ & 0.12 & (0.00, 0.24) \\
  smallpox $\rightarrow$ pertussis & $\beta_{s}^p$ & 0.04 & (-0.04, 0.12) & $\gamma_{s}^p$ & 0.11 & (-0.02, 0.23)\vspace{0.2cm} \\
 
  pertussis $\rightarrow$ measles & $\beta_{p}^m$ & 0.13 & (0.06, 0.20) & $\gamma_{p}^m$ & 0.11 & (-0.01, 0.23) \\
  smallpox $\rightarrow$ measles & $\beta_{s}^m$ & 0.15 & (0.05, 0.24) & $\gamma_{s}^m$ & 0.24 & (0.08, 0.39)\vspace{0.2cm} \\
 
  pertussis $\rightarrow$ smallpox & $\beta_{p}^s$ & 0.05 & (-0.03, 0.13) & $\gamma_{p}^s$ & 0.19 & (0.05, 0.33) \\
  measles $\rightarrow$ smallpox & $\beta_{m}^s$ & 0.21 & (0.12, 0.31) & $\gamma_{m}^s$ & 0.38 & (0.21, 0.54) \\
  \hline
\end{tabular}
\end{table}

For full results of all time and town invariant parameter estimates with their prior and posterior intervals, see Supplementary Table 1 on GitHub.

\subsection*{Model comparison}

While our main interest was studying the past spatio-temporal dynamics of infections and disease associations within and between the diseases, we also examined the necessity and reasonableness of modeling the disease interdependencies and the response aggregation. We compared our model with a corresponding one without the dependencies between the infections by excluding the other diseases as explanatory variables and omitting the correlation between the incidence factors $\tau_t$ in the simpler model. Since the original data contained the numbers of deaths instead of the dichotomous aggregates we used as a response, we also estimated corresponding models with the difference of using the counts as a response and a negative binomial distribution to model them. Formally written,
\begin{equation}
    \label{eq:model_nb}
    y_{i,t}^d \sim \text{NB}\!\left(\!\text{exp}\!\left(\eta_{i,t}^d\!\right)\!\!, \, \text{exp}\!\left(\!\alpha_{\phi^d} + \phi_i^d\right)\right)\!\!,
\end{equation}
where the mean parameter $\eta_{i,t}^d$ is defined as in \autoref{eq:eta}, and the nationwide dispersion parameters $\alpha_{\phi^d}$ and the local dispersion parameters $\phi_i^d$ depend on the response disease. The priors are the same as with the Bernoulli model, with the addition of $\alpha_{\phi^d} \sim \text{N}(0,\, 1)$, $\phi^d \sim \text{N}(0,\, \sigma_{\phi^d})$, and $\sigma_{\phi^d} \sim \text{Gamma}(2,\, 1)$.
From the negative binomial model, we could then compute our quantity of interest, the probability of observing at least one death in a specific town and month, which could be compared with the estimates of the Bernoulli model. Additionally, the briefly aforementioned model without a seasonal component was included in the comparison in the case of both types of responses. This resulted in six different model versions for comparison: dependent diseases, independent diseases, and dependent diseases without a seasonal effect, each for both Bernoulli and negative binomial distributions.

We used the expected log predictive density (ELPD), which measures the goodness of the entire predictive distribution, as a scoring rule for the model comparison \citep{vehtari2017}. The ELPD was estimated via an approximate leave-one-out cross-validation using the R package \textit{loo} \citep{vehtari2023}. We left out one month and town from all the diseases at a time to estimate the ELPD. Models with higher values of ELPD correspond to greater posterior predictive accuracy for predicting new data points compared to models with lower ELPD values.

According to the differences in the ELPDs in \autoref{tab:loo}, the best performing model is the one utilizing Bernoulli distribution and considering the diseases dependent. Omitting the dependencies results in the second-best model, with the difference in ELPD over three times the standard error. As could be expected, omitting the seasonal effect further impairs the model. When it comes to the negative binomial models with counts as responses, the order of the dependent, independent and seasonless models is the same. The Bernoulli models outperform the negative binomial ones in all cases. Overall, we see that directly using the dichotomized data versus modeling the count data has a greater impact than considering the infection dependencies or seasonality in our model. However, even though in terms of predictive performance the differences between different Bernoulli models are relatively small, we used the most complex model in our main analysis, in line with the common Bayesian paradigm of incorporating the uncertainty of the model structure in the model \citep{Vehtari2012}.

\begin{figure}
    \centering
    \includegraphics{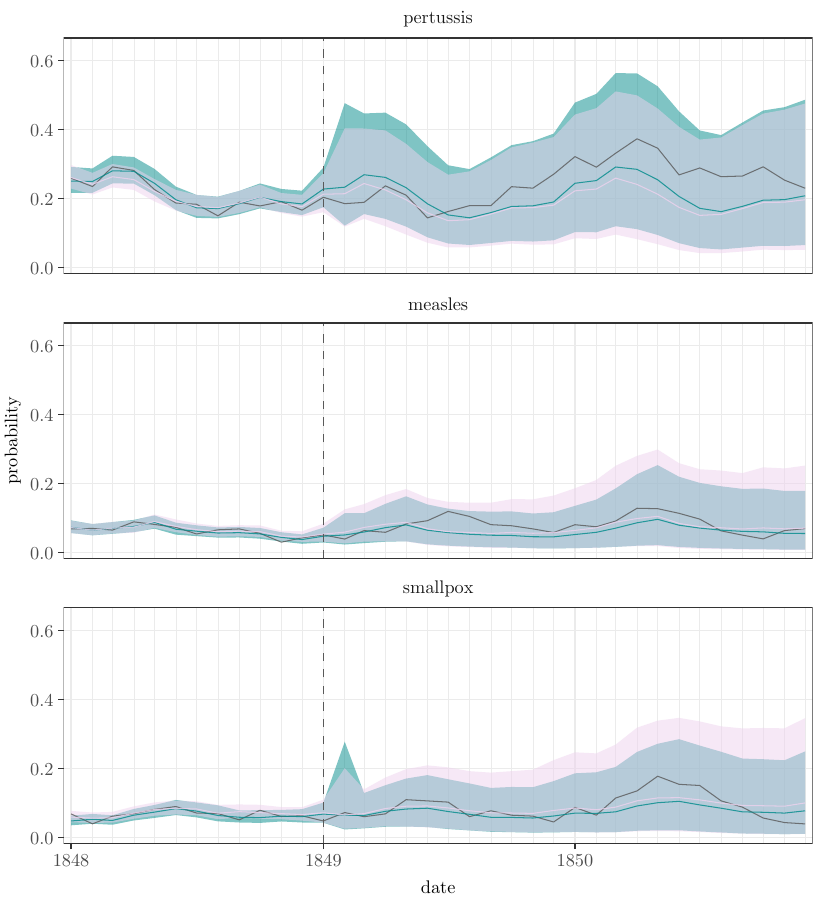}
    \caption{The dark gray lines depict the proportions of the towns where at least one death was observed. The turquoise lines show the posterior means of the corresponding estimates and the shaded areas their $95$\% posterior intervals in the case of the Bernoulli model, whereas the pink lines and areas represent the same values for the negative binomial model. The dotted vertical line indicates the time point after which the model estimates are predicted by the models estimated without the last two years' data.}
    \label{fig:time_pred}
\end{figure}

For the dependent and independent models we performed additional prediction checks by discarding the last two years of the data and estimating the probabilities for those years. We also calculated the ELPDs considering the removed years, see \autoref{tab:loo}. The modifications without the seasonal effect were not included in this comparison due to their already evident poor performance and the fact that they were originally fitted merely to investigate the importance of the obvious seasonal variation. The posterior means were quite similar in all cases, but the posterior intervals were wider for measles and smallpox in the case of the negative binomial model, as can be seen from \autoref{fig:time_pred}.

\begin{table}[ht]
\centering
\caption{\label{tab:loo}Differences of the ELPDs and the standard errors of the ELPD differences for the leave-one-out cross-validation. The values in the first two columns are computed over all the years for the models estimated with the full data, and the last two columns are the values calculated over the last two years for the models discarding those years while estimating the models.}
\begin{tabular}{lrrrr}
  \hline
  & \multicolumn{2}{c}{\hspace{-0.2cm}full data} & \multicolumn{2}{c}{last two years} \\
 & $\textrm{ELPD}_\textrm{diff}$ & $SE_\textrm{diff}$ & $\textrm{ELPD}_\textrm{diff}$ & $SE_\textrm{diff}$ \\ 
  \hline
  Bernoulli dependent & 0.00 & 0.00 & 0.00 & 0.00 \\ 
  Bernoulli independent & -36.71 & 11.75 & -46.84 & 4.21 \\ 
  Bernoulli dependent, without season & -80.50 & 13.08 & - & - \\ 
  Negative binomial dependent & -353.88 & 43.75 & -381.38 & 16.70 \\ 
  Negative binomial independent & -400.59 & 45.01 & -490.50 & 18.90 \\ 
  Negative binomial dependent, without season & -437.72 & 45.98 & - & - \\ 
   \hline
\end{tabular}
\end{table}

Overall, the results of the Bernoulli, as well as the negative binomial, models seem to indicate similar interdependencies between the diseases. In the case of the negative binomial model, the estimates of all the time and town invariant parameters with their prior and posterior intervals are in Supplementary Table 2 on GitHub. Also figures corresponding the ones representing the results of the Bernoulli model are available in GitHub (Supplementary Figures 3-9).

\section*{Discussion}

We developed a Bayesian model to explore the spatio-temporal dynamics and co-dynamics of three childhood infections---measles, smallpox and pertussis---in pre-healthcare Finland (1820--1850). The main novelties of the approach are, firstly, the consideration of both the spatial and temporal aspects simultaneously, and, secondly, considering the connections not only within but also between the three diseases. Furthermore, our dataset is substantially different in comparison to the corresponding previous epidemiological literature. Instead of data regarding large cities or being pooled over countries, we exploited records from sparsely populated pre-healthcare Finland, where $1.2$--$1.6$ million inhabitants were spread over vast areas in hundreds of small towns. When it comes to the explanatory elements of the model, they all capture different features. According to our results, all the components are meaningful and statistically distinctive, and the inclusion of the possibility of dependencies between the diseases leads to a model describing the data better than one merely assuming independent diseases. The data and the model framework are available on GitHub, providing a template for other researchers.

Based on our results, the main components explaining the variation in the probabilities of observing at least one death caused by pertussis, measles or smallpox are the nationwide incidence factors with their local adjustments. The estimated incidence factors follow the temporal behavior of the observed data, and the regional adjustments resemble the spatial patterns of the data (Figures \ref{fig:series} and \ref{fig:tau}, and Figures \ref{fig:maps} and \ref{fig:lambda}).

Measured by pairwise correlations of the incidence factors, a distinctive positive co-occurrence of measles and pertussis was discovered. Previous research has found positive, negative and inconsistent co-occurrences of these infections, e.g. \citet{rohani2003}, \citet{coleman2015}, and \citet{noori2019}. We also found a notable connection between measles and smallpox with a model without the seasonal component, but this correlation is not present in the full model including the seasonality. This indicates that their dynamics follow a similar, seasonal pace. Overall, the seasonal effect is visible among all the diseases. In addition to the nationwide incidence level, the seasonality increases the mortality during the first half of the year, depending on the disease, see \autoref{fig:season}. Seasonalities could reflect increased transmission during social gatherings, or they could be due to environmental and climatic drivers \citep{metcalf2009, metcalf2017}. The work of \citet{briga2021}, based on selected data covering longer periods, indicates that of the infections of pertussis, measles and smallpox only pertussis was linked with new year and Easter in Finland in the 18th and 19th centuries.

Furthermore, lagged dependencies within and between the infections were discovered as positive temporal and spatial effects of the explanatory variables. Recorded deaths in the focal town and in its adjacent towns in the previous month increased the risk of dying of the same disease. Between the infections these effects were notably smaller (\autoref{tab:coefs}). It should be noticed that the coefficients reflecting the effect of the history of the focal town and its neighbors are not directly comparable, as the value of the focal covariate is either $0$ or $1$ but the neighborhood covariate is a proportion between $0$ and $1$.

According to the results, the risk of succumbing to pertussis, measles and smallpox was mediated by occurrences of the other infections in the area. All these three diseases tended to increase the mortality related to the two other diseases, as all the pairwise interaction parameter estimates are positive. This might be due to general immunosuppression or to decreased condition following the previous infection. The strongest associations were found between measles and pertussis, and measles and smallpox. The possibility that pertussis is driven by immunosuppressive effects of measles as suggested by \citet{coleman2015} and \citet{noori2019} implies that the risk of dying of pertussis is increased by a recent measles infection. This is also supported by findings of \citet{mina2015} showing that measles vaccination, by preventing measles-associated immune memory loss, decreases risk of other infections. Our observations (see \autoref{tab:coefs}) are aligned with these results. However, also a reverse connection was recovered: the recorded deaths caused by pertussis in the same town during the previous month increased the risk of observing one or more measles induced deaths almost equally. A stronger lagged effect was discovered between measles and smallpox. Also these interactions were found to act in both directions.

To gain further insights into the specific effects of immunosuppression and impaired health conditions, longer lags than the one month we used here should likely be employed. Unfortunately, our data do not suffice for identifying such effects as accounting longer histories or using finer timescale is challenging due to the missing data and the relative rarity of the deaths. Also population immunity would be an important thing to consider, but without proper population size estimates it remains a topic for future work.

The observed spatially varying local risks of at least one death due to pertussis, measles or smallpox may arise from the closeness of potential sources of infection, differences in cultural, housing or nutritional circumstances, or even genetics \citep{honkola2018, voutilainen2017, kerminen2017}. As can be seen from \autoref{fig:maps}, the probabilities of detecting one or more deaths caused by pertussis and smallpox were greater in the eastern parts of Finland, whereas measles was clearly an infection emphasized in the southern parts, being in concordance what was suggested by \citet{pitkanen1989} and \citet{ketola2021}.

When it comes to the long term temporal behavior of the infections, it seems that epidemics in small populations, consisting of sparse metapopulations of tiny towns, might be dominated by reintroductions and fade outs rather than by endemic dynamics more typical in densely populated cities and countries \citep{keeling1997, grenfell1997, rohani2003, ketola2021}. In \citet{briga2022} epidemics were found to reoccur in cycles of roughly four years in the 18th and 19th centuries in chosen Finnish towns with the highest quality data. The length and phase of such patterns are likely to vary due to annual and geographical differences in seasons, making them challenging to estimate from our scarce data. Our study covering 31 years did not reveal any long term nationwide periodicities.

We modeled the deaths caused by measles, smallpox and pertussis via a binary Bernoulli distribution, where value $1$ denotes that there was at least one reported death given the disease, town and month, and $0$ for no reported deaths. This approach, while sacrificing some detail, allowed us to capture the broad trends and patterns in the data, and to make meaningful inferences about the spatio-temporal co-dynamics of these diseases. In contrast to the generally held view that dichotomizing data should be avoided, in our case directly modeling binary presence-absence data seemed to be beneficial compared to modeling observed death counts, potentially due to accuracy issues in the actual counts. However, both approaches led to identical main conclusions. Our model comparisons also exemplified how our approach is applicable to other kinds of responses than Bernoulli variables.

We accounted for spatial dependencies using explanatory variables based on a neighborhood structure defined by a shared border between two towns. To model and quantify the evident epidemiological transmission dynamics, we included neighbor effects enabling the situation in the adjacent towns in the previous month to affect the probability to observe one or more deaths in the focal town. Our choice of neighborhood is straightforward, omitting the actual intensity of communication between the neighboring towns, hence possibly shrinking or magnifying the true dynamics of the infections. If there were more detailed data or complementary information about the social connections, other definitions for neighborhood, even with an appropriate weighing mechanism, could be employed. We tried to consider each pair of neighbors individually, but the information in the data was not sufficient for model identifiability, owing to the rarity of cases in neighboring towns. Naturally, including alternative appropriate and available covariates as explanatory variables is possible as well. The general spatio-temporal model developed for the purpose of exploring the dynamics and co-dynamics particularly in the case of sparse and scarce data is applicable to other corresponding datasets, for example, based on the historical parish records from other Nordic countries, or data on modern day rural areas.

\section*{Acknowledgements} 
The authors wish to acknowledge CSC – IT Center for Science, Finland, for computational resources. The authors also thank Virpi Lummaa.

\bibliography{bibliography}

\end{document}